\documentclass[twoside, 12pt]{report}
\usepackage[english]{babel}
\usepackage{epsfig}
\usepackage{psfig}
\usepackage{amssymb,amsmath,amsfonts,soul,amsthm,enumerate}
\usepackage{graphicx}
\usepackage[cp1251]{inputenc}
\usepackage[T2A]{fontenc}
\usepackage{mathrsfs}
\usepackage{algorithm2e}
\def\@evenfoot{}
\def\@oddfoot{}

\begin{document}
\def\@evenhead{\vbox{\hbox to \textwidth{\thepage\leftmark}\strut\newline\hrule}}

\def\@oddhead{\raisebox{0pt}[\headheight][0pt]{%
\vbox{\hbox to \textwidth{\rightmark\thepage\strut}\hrule}}}

\newpage
\normalsize
\def\bibname{\vspace*{-30mm}{\centerline{\normalsize References}}}
\thispagestyle{empty}
\begin{flushleft}
{\small {\bf Russian University of Transport.\\Department of Computing Systems, \\Networks and Information Security.}\\
}
\end{flushleft}
\vskip 5 mm

\centerline{\bf METHOD FOR DETERMINING THE}
\centerline{\bf ACCELERATION OF A PARALLEL SPECIALISED}
\centerline{\bf COMPUTER SYSTEM BASED ON AMDAHL'S LAW}
 
\vskip 0.3cm
\centerline{\bf Aleksandr S. Filipchenko}

\vskip 0.3cm
\noindent{\bf Abstract}
{\small The modification of Amdahl's law for the case of increment of processor elements in a computer system is considered. The coefficient $k$ linking accelerations of parallel and parallel specialized computer systems is determined. The limiting values of the coefficient are investigated and its theoretical maximum is calculated. It is proved that $k$ > 1 for any positive increment of processor elements. The obtained formulas are combined into a single method allowing to determine the maximum theoretical acceleration of a parallel specialized computer system in comparison with the acceleration of a minimal parallel computer system. The method is tested on Apriori, k-nearest neighbors, CDF 9/7, fast Fourier transform and naive Bayesian classifier algorithms.
}
\vskip 0.3cm

\setcounter{figure}{0}

\renewcommand{\thesection}{\large 1}

\section{\large Introduction}

As we know, development of a parallel specialized computer system is a labor-intensive task.
That's why before diving into this process we need to answer a number of questions.
Namely, will a parallel specialized computer system cope with processing of the chosen class of tasks better than a usual one?
And if yes, how noticeable will this effect be? Let's find out.

\renewcommand{\thesection}{\large 2}
\section{\large Notations and definitions}

One of the most widespread and time-tested methods of estimating performance of parallel computing systems is application of Amdahl's law~\cite{Amdahl}.
For an ordinary parallel computing system the acceleration by Amdahl's law is defined by formula~\ref{f1}.
\begin{gather}\label{f1}
S = \frac{1}{f+(1-f)P^{-1}},
\end{gather}
where $S$ is the acceleration due to parallelization, $f$ is the share of non-parallelized computations in the total volume and $P$ is the number of processors~\cite{Amdahl-Cheremisinov}.

Based on this law, we will represent the acceleration realised by a parallel specialised computer system as an increment of the number of processors in the expression~\ref{f2}.
\begin{gather}\label{f2}
S^\prime = \frac{1}{f+(1-f)(P + \Delta P)^{-1}},
\end{gather}
where $S^\prime$ is the acceleration of a parallel specialised computing system.

Let $S$ and $S^\prime$ be related to each other by some proportionality coefficient $k$: $S^\prime = kS$. Let's substitute expressions from expressions~\ref{f1} and~\ref{f2} into this equality. We get formula~\ref{f3}.
\begin{gather}\label{f3}
\frac{1}{f+(1-f)P^{-1}} = \frac{k}{f+(1-f)(P +\Delta P)^{-1}}.
\end{gather}

Thus, the coefficient $k$ can be represented in the following form~\ref{f4}.
\begin{gather}\label{f4}
k = \frac{f + (1-f)P^{-1}}{f + (1-f)(P + \Delta P)^{-1}}.
\end{gather}

\renewcommand{\thesection}{\large 3}
\section{\large Proof of the superior acceleration of a specialised computing system on a conventional one}

Assume that this coefficient is greater than one, and reflect the solution of the corresponding inequality in expressions~\ref{f5},~\ref{f6},~\ref{f7} and~\ref{f8}.
\begin{gather}
\frac{f + (1-f)P^{-1}}{f + (1-f)(P + \Delta P)^{-1}} > 1\label{f5} \\
f + \frac{1-f}{P} > f + \frac{1-f}{P + \Delta P}\label{f6} \\
\frac{1-f}{P} > \frac{1-f}{P + \Delta P}\label{f7} \\
P + \Delta P > P\label{f8}
\end{gather}

Since the expression $P + \Delta P > P$ is true for any $\Delta P > 0$, then $S^\prime > S$, which was required to prove.

\renewcommand{\thesection}{\large 4}
\section{\large The theoretical maximum of $k$}

Assuming that the variable $f$ for a particular algorithm is given and the number of processors $P$ in the minimal parallel system is two, the maximum theoretical value of $k$ for a particular algorithm can be found using the limit at $\Delta P$ tending to infinity. This fact is reflected in formula~\ref{f9}.

\begin{gather}\label{f9}
max(k) = \lim_{\Delta P \to \infty} \frac{f + (1-f)P^{-1}}{f + (1-f)(P + \Delta P)^{-1}} = \frac{f(P-1)+1}{Pf}=\frac{f+1}{2f}
\end{gather}

Consequently, the coefficient of parallel processing acceleration in a specialised computer system depends only on the algorithm being implemented, namely on the degree of its parallelizability.
That is, in order to rationalise the development of a parallel special-purpose computer system, it is enough to analyse algorithms from the corresponding class of tasks, find the maximum $k$ for each of them and take the algorithm with the highest $k$ as a basis.

\renewcommand{\thesection}{\large 5}
\section{\large Approval of the method on a number of Big Data algorithms}

The following big data algorithms have been widely used in the transport industry:
\begin{itemize}
\item Apriori algorithm~\cite{Pereslegin, Schabe};
\item k-nearest neighbours algorithm~\cite{YinG};
\item wavelet transform~\cite{Fourier};
\item Fourier transform~\cite{Fourier};
\item naive Bayesian classifier~\cite{Croatia}.
\end{itemize}

To validate our method, we need to analyse parallel implementations of these algorithms in order to find the percentage of separable steps to the total amount of computation.

\subsection{Apriori algorithm}\label{t-apriori}
The Apriori algorithm searches for associative rules and is applied to databases with a huge number of transactions.
The algorithm consists of two main stages: generation and counting.
The generation phase scans the database to generate a set of candidate frequent subsets.
The counting stage discards candidates whose number of transactions is less than the minimum value.
Thus, the algorithm iteratively returns only the most frequently occurring sets in the data set.
The Apriori algorithm is formalised by the following pseudo code~\cite{Apriori}.

\RestyleAlgo{ruled} 
\begin{algorithm}
\caption{Apriori algorithm pseudo code}
\KwData{$T$, $\varepsilon$}
\Begin {
  $L_{1} \gets \{large\ 1 - itemsets\}$\;
  $k \gets 2$\;
  \While{$L_{k-1} \neq \varnothing$} {
    $C_{k} \gets Gen(L_{k-1})$\;
    \For{transactions $t \in T$} {
      $C_{t} \gets \{c \in C_k : c \subseteq t\}$\;
      \For{candidates $c \in C_t$} {
        $count[c] \gets count[c] + 1$\;
      }
    }
    $L_{k} \gets \{c \in C_k : count[c] \geqslant \varepsilon\}$\;
    $k \gets k+1$\;
  }
}
\KwResult{$\bigcup_{k} L_{k}$}
\end{algorithm}

Thus, the Apriori algorithm consists of ten main steps.
We take the computational volume of these ten operations as 100\%.
For simplicity, we assume that each stage includes a computational volume equal to 10\%.

Stage $L_{1} \gets \{large\ 1 - itemsets\}$ is perfectly parallelised: nothing prevents us from performing a parallel search of the database for several items at once.
Each processor can look for its own itemset independently of each other.
In this case, the forwarding procedures $k \gets 2$, $count[c] \gets count[c] + 1$ and $k \gets k+1$ are too elementary to be parallelised.
This is a sequential part of the algorithm.
The procedure $C_{k} \gets Gen(L_{k-1})$ glues the elements into sets, forming their combinations.
It can also be performed in parallel, for example, by gluing elements $I1, I2, I3, I4$ into pairs, resulting in arrays $\{I1, I2\}, \{I1, I3\}, \{I1, I4\}, \{I2, I3\}, \{I2, I4\}, \{I3, I4\}$.
The same is true for the inverse operation of subset selection $C_{t} \gets \{c \in C_k : c \subseteq t\}$.
Operation $L_{k} \gets \{c \in C_k : count[c] \geqslant \varepsilon\}$, which discards candidates with insufficient frequency from the list $L$, is also easily paralleled into $c$ comparator threads.

It remains to deal with loops.
Usually loops are well-parallelised program structures.
This is due to the fact that it is often possible to execute a number of loop iterations in parallel.
The for loops in this algorithm are definitely such loops.
Each new iteration in them simply loops through the arrays available to them.
There is no rigid connection between array data there, and therefore there are no obstacles for parallel processing either.
The main while loop is quite different.
The correctness of its operation is directly related to stage $L_{k} \gets \{c \in C_k : count[c] \geqslant \varepsilon\}$, which is formed throughout the cycle.
Parallel operation will break the logic of this loop and the whole programme, so the iterations of this loop must be executed strictly sequentially.
Consequently, we can conclude that the sequential part in the Apriori algorithm takes 40\%.
Then the maximum of the coefficient $k$ in this case can be represented by the expression~\ref{f10}.

\begin{gather}\label{f10}
max(k_{Apriori}) = \frac{f+1}{2f} = 1.75
\end{gather}

\subsection{k-nearest neighbors algorithm}\label{t-knn}
The k-nearest neighbours algorithm classifies objects in the informative feature space $T$.
An object belongs to the class which has more neighbours.
The number of neighbours k is defined as an odd number, usually 3, 5 or 7.
The distance to the neighbours is determined by the Pythagorean theorem.
The feature space $T$ and the feature vector of the unidentified object $\vec{x}$ are passed to the function of the k-nearest neighbours algorithm.
The function returns the class to which the object $x$ belongs.

\begin{algorithm}
\caption{k-nearest neighbours algorithm pseudo code}
\KwData{$T$, $\vec{x}$}
\Begin {
  $k \gets 2m+1, m \in \mathbb Z$\;
  \For{$i \gets 1$ \KwTo $n(T)$} {
    $d_{i} \gets \sqrt{\sum\limits_{j=1}^{n(\vec{x})}(x_{j}-T_{ij})^2}$\;
  }
  $sort(d)$\;
  $votes \gets \varnothing$\;
  \For{$i \gets 1$ \KwTo $k$} {
    \lIf{$neighbor_{i} \in A$}{$votes_{i} \gets 1$}
  }
  \lIf{$\sum\limits_{i=1}^{k} votes_{i} \geqslant k/2$} {\textbf{Result:} $x \in A$}
}
\KwResult{$x \notin A$}
\end{algorithm}

Let us repeat the same way of splitting the algorithm into parallel and sequential parts by analysing its pseudocode.
This algorithm consists of nine main steps.
Assignment operations $k \gets 2m+1, m \in \mathbb Z$,   $votes \gets \varnothing$ and $votes_{i} \gets 1$ are sequential.
The first for loop can be parallelised, since nothing prevents us from calculating distances to several points at once.
The same is true for the calculation of the Euclidean metric.
Sorting step serves only for sorting distances from an object to its neighbours, which will print the distances and indices of the nearest ones to the top of the array.
This means that the sorting algorithm can be used arbitrarily, including multi-threaded implementations.
For second loop and $neighbor_{i} \in A$ there are also no obstacles to parallelisation.
At the same time, in the branching procedure compared to $\sum\limits_{i=1}^{k} votes_{i} \geqslant k/2$ it is rather problematic to find candidates for parallelisation, so this step will be performed sequentially. It turns out that only 5 out of 9 operations are parallelised.
Then the maximum of the coefficient $k$ in this case can be represented by the expression~\ref{f11}.

\begin{gather}\label{f11}
max(k_{KNN}) = \frac{f+1}{2f} = 1.625
\end{gather}

\subsection{Cohen-Daubechies-Feauveau wavelet algorithm}\label{wavelet}
For this algorithm, the pseudocode analysis can be omitted since the counting of separable, i.e. parallelisable, and non-separable, i.e. sequential, stages is performed in the corresponding publication~\cite{CDF}.
The publication shows that the CDF 9/7 implementation includes a total of 21 stages, 14 of which are separable.
Hence, the amount of sequential and parallel computations for the <<Cohen---Daubechies---Feauveau>> wavelet can be represented as 1/3.
Then the maximum of the coefficient $k$ in this case can be represented by the expression~\ref{f12}.

\begin{gather}\label{f12}
max(k_{CDF 9/7}) = \frac{f+1}{2f} = 2
\end{gather}

\subsection{Fast Fourier transform algorithm}\label{FFTalg}
The discrete Fourier transform algorithm and its fast implementation are difficult to analyse because of their complexity.
Thus, the lower bound of complexity of fast Fourier transform algorithms has not been found yet.
That is, it is not known whether it can be faster than $O(n \log n)$.
In this regard, the experimental method remains an available possibility for studying Fourier transform algorithms.
The essence of the method is to develop a program that performs this transformation and study its execution by profiling.

As we know, the Cooley---Tukey algorithm that performs the fast Fourier transform consists of three main steps:
\begin{itemize}
\item calculation of the rotation coefficient;
\item the butterfly operation;
\item multiplication of complex numbers.
\end{itemize}

Using the <<DS-5 Streamline>> profiling tool, it was determined how much per cent of computational time each of these sections required~\cite{FFT}.
The calculation of the rotation factor takes 25.4\% of the time, the butterfly operation takes 21.8\%, and the multiplication of complex numbers takes 18.2\%.
The butterfly operation consists of three nested cycles.
In this case, the outer loop, unlike the two inner ones, is not subject to parallelisation because it depends on the result of the previous iteration~\cite{FFT}.
That is, the amount of calculations in the algorithm can be represented as expression~\ref{f13},~\ref{f14}.

\begin{gather}
V_{FFT} = 100\% = f + p \label{f13} \\
f = 100\% - p = 100\% - (25.4\% + 21.8\% \frac{2}{3} + 18.2\%) = 41.8\overline{6}\% \label{f14}
\end{gather}

Then the acceleration coefficient from the increment of processor elements according to Amdahl's law will take the approximate value by formula~\ref{f15}.

\begin{gather}\label{f15}
max(k_{FFT}) = \frac{f+1}{2f} \approx 1.69
\end{gather}

\subsection{Naive Bayesian classifier}\label{NBCalg}
A parallel algorithm for a Naive Bayesian classifier is presented as a flowchart in <<A novel parallel implementation of Naive Bayesian classifier for Big Data>>~\cite{NBC}.
The flowchart shows that the algorithm consists of seven steps.
At that, the last stage consists of three operations:
\begin{itemize}
\item global table search;
\item classification in the global table or by multiplication of probabilities;
\item writing to the global table or to the classifier.
\end{itemize}
That is a total of nine operations, four of which are parallelised and five of which are not.
The amount of non-parallelised computations is represented as a ratio of 5 to 9.
Then the maximum of the coefficient $k$ in this case can be represented by the expression~\ref{f16}.

\begin{gather}\label{f16}
max(k_{NBC}) = \frac{f+1}{2f} = 1.4
\end{gather}

\renewcommand{\thesection}{\large 6}
\section{\large Analysing the results of approbation}

\begin{figure}[tbp]
\begin{center}
%\begin{psfrags}
%\small
%\includegraphics{FilipchenkoA_fig1.eps}
%\end{psfrags}
\epsfig{file=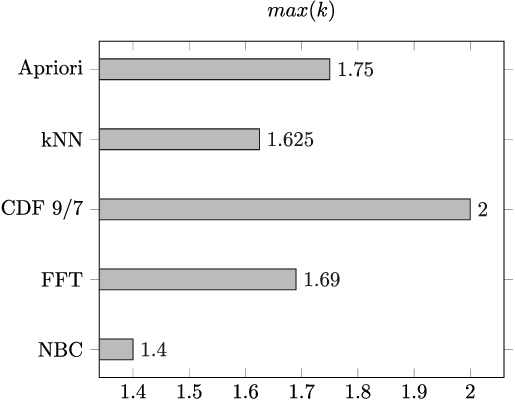}
\end{center}
\caption{Comparison of specialised acceleration coefficients of Big Data algorithms}
\label{fig}
\end{figure}

Let us present the obtained estimates for all the considered algorithms in the form of a bar chart shown in figure~\ref{fig}.
It can be seen that the most promising algorithm for specialisation of a parallel computing system out of the considered ones is the wavelet transform, specifically its implementation CFD 9/7.

\section*{\large Acknowledgement}
The author would like to thank the Head of the Department <<Computer Systems, Networks and Information Security>> of the Russian University of Transport, PhD, Associate Professor Boris V. Zhelenkov for the scientific guidance of the article.

%%%%%%%%%%%%%%%%%%%%%%%%%%%%%%%%%%%%%%%%%%%%%%%%%%%%%%%%%%%%%%%%%%%%%%%%%%%
%\thispagestyle{headings}

\begin{flushleft}
Aleksandr S. Filipchenko,\\
Russian University of Transport, \\
Russia, 127994, Moscow, Obraztsova str., 9-9,\\
Email: {\tt 797439@edu.rut-miit.ru}\\
\end{flushleft}


\begin{thebibliography}{99}
\thispagestyle{myheadings}
%%%%%%%%%%%%%%%%%%%%%%%%%%%%%%%%%%%%%%%%%%%%%%%%%%%%%%%%%%%%%%%%%%%%%%%%%%%
\vspace*{-10mm}
\footnotesize
\bibitem{Amdahl}
{Amdahl G. M.}, {\it Validity of the Single Processor Approach to Achieving Large-Scale Computing Capabilities}, AFIPS Conference Proceedings, 1967, pp. 483--485, doi: 10.1145/1465482.1465560.

\bibitem{Amdahl-Cheremisinov}
{Cheremisinov D. I.}, {\it Amdahl's law and bounds on speedup}, Big Data and Advanced Analytics, no. 6--2 (2020), pp. 295--301.

\bibitem{Pereslegin}
{Pereslegin S. V., Karpov I. O., Khalikov Z. A.}, {\it Two-position quasi-mirror radar of the sea surface: Principles of microwave scattering and possibilities of solving Oceanology problems from space}, Oceanology, vol. 57, no. 5 (2017), pp. 639--647. doi: 10.1134/S0001437017050149.

\bibitem{Schabe}
{Sch\"abe H.}, {\it Autonomous Driving – How to Apply Safety Principles}, Dependability, no. 19 (2019), pp. 21--33. doi: 10.21683/1729-2646-2019-19-3-21-33.

\bibitem{YinG}
{Yin G., Huang Z., Yang L., Ben-Elia E., Xu L., Scheuer B., Liu Y.}, {\it How to quantify the travel ratio of urban public transport at a high spatial resolution? A novel computational framework with geospatial big data}, International Journal of Applied Earth Observation and Geoinformation, vol. 118 (2023), doi: 10.1016/j.jag.2023.103245.

\bibitem{Fourier}
{Popov B. N., Fedorina E. S.}, {\it Using of methods for analysis and processing data to an information flows for objects of water transport}, Vestnik gosudarstvennogo universiteta morskogo i rechnogo flota imeni admirala S.O. Makarova, no. 30 (2015), pp. 220--225, doi: 10.21821/2309-5180-2015-7-2-220-225.

\bibitem{Croatia}
{Vidovi\'c K., \v{C}oli\'c P., Vojvodi\'c S., Blavicki A.}, {\it Methodology for public transport mode detection using telecom big data sets: case study in Croatia}, Transportation Research Procedia, no. 64 (2022), pp. 76--83, doi: 10.1016/j.trpro.2022.09.010.

\bibitem{Apriori}
{Agrawal R., Srikant R.}, {\it Fast algorithms for mining association rules}, Proceedings of the 20th International Conference on Very Large Data Bases, 1994, pp. 487--499.

\bibitem{CDF}
{Ba\v{r}ina D., Kula M., Matysek M., Zemc\'ik P.}, {\it Accelerating Discrete Wavelet Transforms on Parallel Architectures}, Journal of WSCG, vol. 25 (2017), pp. 77--85, e-print arXiv:1704.08657v2.

\bibitem{FFT}
{Vincke R., Van L. S., Cordemans P., Peuteman J., Steegmans E., Boydens J.}, {\it Calculating Fast Fourier Transform by using parallel software design patterns}, KU Leuven, 2012, doi: 10.13140/RG.2.2.11694.51522.

\bibitem{NBC}
{Katkar V. D., Kulkarni S. V.}, {\it A novel parallel implementation of Naive Bayesian classifier for Big Data}, International Conference on Green Computing, Communication and Conservation of Energy (ICGCE), 2013, pp. 847--852, doi: 10.1109/ICGCE.2013.6823552.
\end{thebibliography}
\end{document}